\documentclass[12pt]{article}
\usepackage{newtxtext}
\usepackage{amsmath,amssymb,amsthm}
\usepackage{newtxmath} 

\usepackage[pdftex]{graphicx}
\usepackage{xcolor}
\usepackage{fancyhdr}

\title{
Modeling Tennis In-Match Momentum Using Probability Method
}
\author{
Jackson Graves, Daniel X. Guo\footnote{Corresponding author email address: guod@uncw.edu},  Ridge Shepherd, and Alexander Young \\
{\em Department of Mathematics and Statistics} \\
{\em University of North Carolina at Wilmington, NC 28403} \\
}

\date{\empty}

\begin{document}

\maketitle

\begin{abstract}
This paper investigates the Tennis Momentum Model (TMM), which aims to enhance the understanding of match dynamics by integrating key factors such as efficiency, historical scoring probabilities, and real-time scoring data. The model is designed to explore how momentum affects player performance throughout a match and how it might influence overall match outcomes. By leveraging this model, players and coaches could gain valuable insights that may help them adjust their strategies in response to shifting momentum during a match.

To validate the model, it was tested on two tennis matches, revealing its effectiveness in capturing shifts in momentum and correlating these shifts with scoring events. The results showed that the TMM accurately depicted the flow of momentum during matches, highlighting how shifts in momentum are directly linked to changes in scoring as the match progresses.

\thanks{\quad {\bf key words}: iid, Bayes estimation, predictive model, scoring probability, short-term momentum, long-term momentum, efficiency}


\end{abstract}

\section{Introduction}
Sports have the unique ability to bring people together worldwide, deeply impacting the lives of individuals and communities. As a result, extensive research has been conducted on various facets of sports, with one of the most intriguing concepts being momentum. In many sports, momentum is seen as a series of shifting peaks and valleys, influenced by a variety of factors that change throughout the game. While the nature of momentum differs across sports, the core idea remains the same: momentum can significantly alter the trajectory of a match.

This paper focuses on tennis and examines how momentum fluctuates during a match. Tennis, with its unique rhythm and mental demands, provides an interesting lens through which to study these shifts. A key challenge, however, in developing momentum models for sports is their ability to generate efficient, accurate, and timely results, particularly when analyzing the dynamic flow of a match in real time.

Tennis is a sport where two players compete against each other in a neutral setting, with no external influences, such as crowd energy, playing a direct role in momentum shifts. Unlike other team sports where crowd support and emotional peaks can drive momentum, tennis players primarily generate momentum through their own performance. Key factors such as the quality of serves and the number of successful rallies are instrumental in shaping the flow of the match. Serving ability, in particular, has a significant impact on momentum, as a strong serve can put a player in a position of control, while a series of successful rallies can build confidence and shift momentum. Thus, for the purposes of this study, momentum changes are hypothesized to be driven by two primary factors: the effectiveness of a player’s serve and the rally count, both of which reflect the player’s technical and psychological state throughout the match.

To model the dynamic nature of momentum in tennis, we introduce a non-iid (non-identically and independently distributed) assumption, which reflects the fact that the probability of scoring on any given rally is not consistent. Rather, it is influenced by the momentum established by previous points. In other words, the outcome of a rally is affected by the psychological and strategic impact of the preceding points, creating a dynamic shift in the probabilities of winning subsequent points.

To capture this, we incorporate both historical and instant scoring probabilities into the Tennis Momentum Model (TMM). The instant scoring probability reflects the immediate advantage gained from winning the most recent point, acknowledging the psychological impact of momentum in the short term. On the other hand, the historical scoring probability accounts for the long-term accumulation of momentum, representing how a series of successful points or games can progressively shift the overall match dynamics. Together, these probabilities enable the model to more accurately reflect the complex and evolving nature of momentum in tennis, showing how past points can influence both immediate and future performance.

To incorporate the evolving nature of momentum, we use empirical Bayes estimation to model the rolling serve percentage of a player as the match progresses. This statistical technique allows us to dynamically estimate a player's serve percentage, adjusting as the match unfolds. The model hypothesizes that a higher serve percentage contributes to increased momentum, which in turn improves the likelihood of winning the match.

Furthermore, we integrate historical data for each player to understand their average serve percentage and expected points played per game. This historical context is critical, as it enables the model to adjust based on a player's individual performance metrics, accounting for the player’s previous statistics. By doing so, the model tailors its predictions, offering a more personalized and accurate assessment of momentum changes. The combination of empirical Bayes estimation and historical data allows the model to provide a nuanced view of how momentum accumulates over time, reflecting both short-term fluctuations in serve performance and long-term trends based on the player’s typical performance patterns.

In our model, efficiency is computed by focusing on key factors that directly impact a player's performance and momentum during a match. These factors include consecutive points won, aces served, the number of back-and-forth rallies, and double faults. While all of these elements contribute to a player's efficiency, the model was simplified to focus primarily on rally length and aces. Rally length is indicative of a player’s ability to sustain control during exchanges, while aces reflect the immediate impact of a player’s serve in gaining an advantage.

One common issue with many sports models is their ability to produce efficient and accurate representations of data in a time-critical and cost-efficient manner \cite{Volleyball}. This challenge is amplified in individual sports like tennis, which contrasts sharply with team sports such as football and basketball. In team sports, where multiple players collaborate, models can focus on group dynamics and strategies. However, tennis, particularly in its singles format, features a one-on-one competition where each player is solely responsible for their performance, making momentum shifts much more frequent and abrupt \cite{grandslam}.

Many existing momentum models, such as the Hot Hand Model \cite{football}, tend to focus on a single player at a time, assessing when they are considered “hot” or in a streak of good performance. While this approach works in team sports, where players may feed off each other, it doesn't fully capture the nature of momentum in singles tennis, where each player's success or failure is entirely dependent on their own performance. In singles tennis, momentum changes are extremely likely, often occurring unexpectedly due to various in-match factors, such as serve efficiency, rally length, and psychological pressure. Thus, a tennis match is characterized by continuous, dynamic shifts in momentum, influenced by a wide range of variables, making the modeling process both more complex and critical to understanding match outcomes.

Momentum in tennis can be defined as a measure of a player's advantage, influenced by previous events in the match. However, momentum is not the only factor at play. Player tendencies, such as their behavioral patterns and responses to specific in-match situations, also play a crucial role in how a match unfolds. For example, some players may thrive under pressure, while others might struggle when momentum shifts against them.

In addition to player tendencies, historical data significantly enhances the predictive power of a momentum model. By analyzing past performances, the model can account for each player's strengths and weaknesses in different match scenarios. One of the most important factors in tennis is serving. A player with a high serving efficiency can quickly gain momentum, as they are more likely to win points decisively, keeping rallies short and often one-sided. This ability to dominate through serving is a critical factor in the development of momentum.

Therefore, a model designed to track momentum swings in tennis must consider not only a player’s serving efficiency but also their ability to finish rallies promptly. When a player can consistently close out points with strong serving and timely rally finishes, they are more likely to maintain control of the match, creating sustained momentum.

In tennis, matches typically begin with predetermined favorites and underdogs based on various factors like player ranking and past performance. However, as the match progresses, these odds can shift rapidly due to momentum changes. The goal of the Tennis Momentum Model (TMM) is to predict these shifts within a single match, providing coaches with a tool to help their players navigate and control momentum swings effectively.

\section{Methodology}

Previous research has demonstrated the importance of momentum as a predictive factor in sports outcomes. For example, deWinter et al. \cite{football} found that incorporating momentum into models significantly improved the ability to predict the outcomes of American football games. Their work highlighted how certain in-game events, such as turnovers and scoring, can shift the momentum and ultimately influence the game's result.

Similarly, Bayrack and Gifford \cite{regression} achieved success by integrating momentum-related events—such as offensive turnovers, third-down conversions, and scoring—into a logistic regression model to predict the outcome of football games. These studies suggest that specific in-game factors, when modeled effectively, can capture the momentum shifts that are crucial in determining match outcomes.

Drawing inspiration from these models, our Tennis Momentum Model (TMM) aims to similarly leverage key in-match events—like serving efficiency and rally control—to track and predict momentum shifts in tennis matches. By identifying and quantifying these momentum events, the TMM seeks to provide insights into how momentum impacts a player’s performance and the overall trajectory of a match.

\subsection{Historical Scoring Probability}
We begin by assuming that, initially, the probability of scoring in a rally is independent and identically distributed (iid), meaning that it remains constant across all rallies. This assumption is a starting point, reflecting the idea that, without accounting for momentum or other in-game factors, the likelihood of winning a point remains the same throughout the match.

For the player serving, we calculate their historical probability $p_{hist}$  of winning a point on serve based on their previous performance. This probability is determined by the formula:
\begin{equation}
p_{hist} = \frac{\text{Total Points Won on Serve}}{\text{Total Serve Attempts}}
\end{equation}
This measure reflects the player's historical success rate on serve, providing a baseline probability of winning a point while serving. By tracking this value, we can assess the player’s serving efficiency, which is a crucial factor in understanding how momentum develops during the match.

If player $p$ is not serving then their probability of scoring would be
\begin{equation*}
    p_{hist} = 1 - q_{hist},
\end{equation*}
where
$$
q_{hist} = \frac{\text{Total Points Won on Serve}}{\text{Total Serve Attempts}}
$$
for opposing player $q$.

The historical scoring probability $p_{hist}$ can serve as a valuable tool for predicting the outcome of a match, especially when two players have previously faced each other. By analyzing their past encounters, we can estimate the probability of a player winning a game based on their historical performance against that opponent. This is particularly useful in cases where detailed historical data exists, as the more data available, the more reliable and accurate the prediction becomes.

For instance, if Player A has consistently won a high percentage of points on serve against Player B in previous matches, we can use this information to predict that Player A is more likely to perform similarly in future encounters, especially if both players' forms have remained consistent.

As the volume of historical data increases, our predictions become more robust, accounting not just for individual player tendencies but also for patterns that may emerge from head-to-head matchups. This offers an opportunity for coaches and analysts to develop more informed strategies based on past results and anticipated momentum shifts, thereby enhancing the overall effectiveness of match preparation.

\subsection{Instant Scoring Probability}
An additional assumption in our model formulation is the existence of long-term momentum. Long-term momentum refers to the cumulative advantage (or disadvantage) that develops for a player throughout a match. As the match progresses, a player can build momentum, leading to an increased likelihood of winning subsequent points. Conversely, negative long-term momentum represents a growing disadvantage that negatively impacts a player's chances of winning future points.

To model this evolving advantage (or disadvantage), we employ an empirical Bayes estimator to estimate the probability of winning a serve during the course of the match, denoted as $p_{inst}$. This estimator dynamically adjusts the probability based on the player’s performance throughout the match, taking into account the momentum shifts as they occur. By updating the probability of winning a point on serve with each game, the empirical Bayes estimator helps reflect the changing dynamics of momentum, capturing both positive and negative trends as they unfold.

This approach allows the model to integrate the effect of long-term momentum, enabling more accurate predictions of a player's performance as the match progresses. It provides insight into how momentum builds over time, influencing both short-term outcomes (such as individual rallies) and the broader trajectory of the match.
\begin{equation}
p_{inst} = \frac{\text{Total Points Won on Serve within a Match}}{\text{Total Serve Attempts within a Match}}
\end{equation}

With the introduction of the instant scoring probability
$p_{inst}$, which is influenced by in-match performance, we now have a factor that dynamically adjusts to reflect momentum shifts during the match. To integrate this into our model, we first theorized the relationship between the historical scoring probability
$p_{hist}$ and the instant scoring probability
$p_{inst}$ over the course of the match.

At the start of the match, there is no in-match data to inform the probability of scoring, making the historical scoring probability
$p_{hist}$ the best estimator for the player's likelihood of winning a point. This is because, at this stage, previous match data and player tendencies are the most reliable indicators of performance.

However, as the match progresses, it is expected that
$p_{inst}$  will become a much more valuable estimator of the probability of scoring. The increasing amount of in-match data, such as recent performance and momentum changes, will make
$p_{hist}$  more reflective of the current dynamics of the match. Conversely, as momentum shifts and player performance evolves, the historical scoring probability
$p_{hist}$  will gradually become a weaker predictor, as it is less sensitive to real-time changes in the match.

This transition allows the model to dynamically update its predictions as momentum builds or wanes, providing more accurate insights into the flow of the match. In essence, the model adapts by weighting
$p_{inst}$  more heavily as in-match performance begins to dominate, while still using the historical probability
$p_{hist}$   as a baseline early on.

\subsection{Efficiency}
We considered several potential parameters that could influence a player's efficiency during the match. These included factors such as consecutive points won in a row, scoring on an ace, the number of back-and-forth rallies, and double faults. Each of these elements contributes to the overall efficiency a player demonstrates during the match.

However, after careful consideration, we simplified the model by focusing on just two key parameters: rally length and aces. The rationale behind this simplification is based on the idea that shorter rallies often indicate greater player efficiency, as they require fewer exchanges to win a point. If a player consistently wins points in short rallies, it suggests they are exerting pressure on their opponent and demonstrating superior control over the flow of the match. On the other hand, aces, which are points won outright without the opponent touching the ball, are the ultimate demonstration of serve dominance and efficiency.

The efficiency of a player is defined as a function of these key parameters,
\begin{equation}
    E = 2 - \sum_{n=0}^k \frac{1}{r^n},
\end{equation}
where $r (r>1)$ is a constant that is determined from historical data and k is the number of rallies in a point. For example, when there are 3 rallies in a point E will be given by,
\begin{equation}
    E = 1 - \frac{1}{r} - \frac{1}{r^2} - \frac{1}{r^3}.
\end{equation}
Note that when k = 0 we are left with
\begin{equation}
    E = 2 - \frac{1}{r^0} = 2 - 1 = 1.
\end{equation}

In figure 1 it can be seen how player efficiency will vary throughout a single set.

\begin{figure}[h!]
    \centering
    \includegraphics[width=0.75\linewidth]{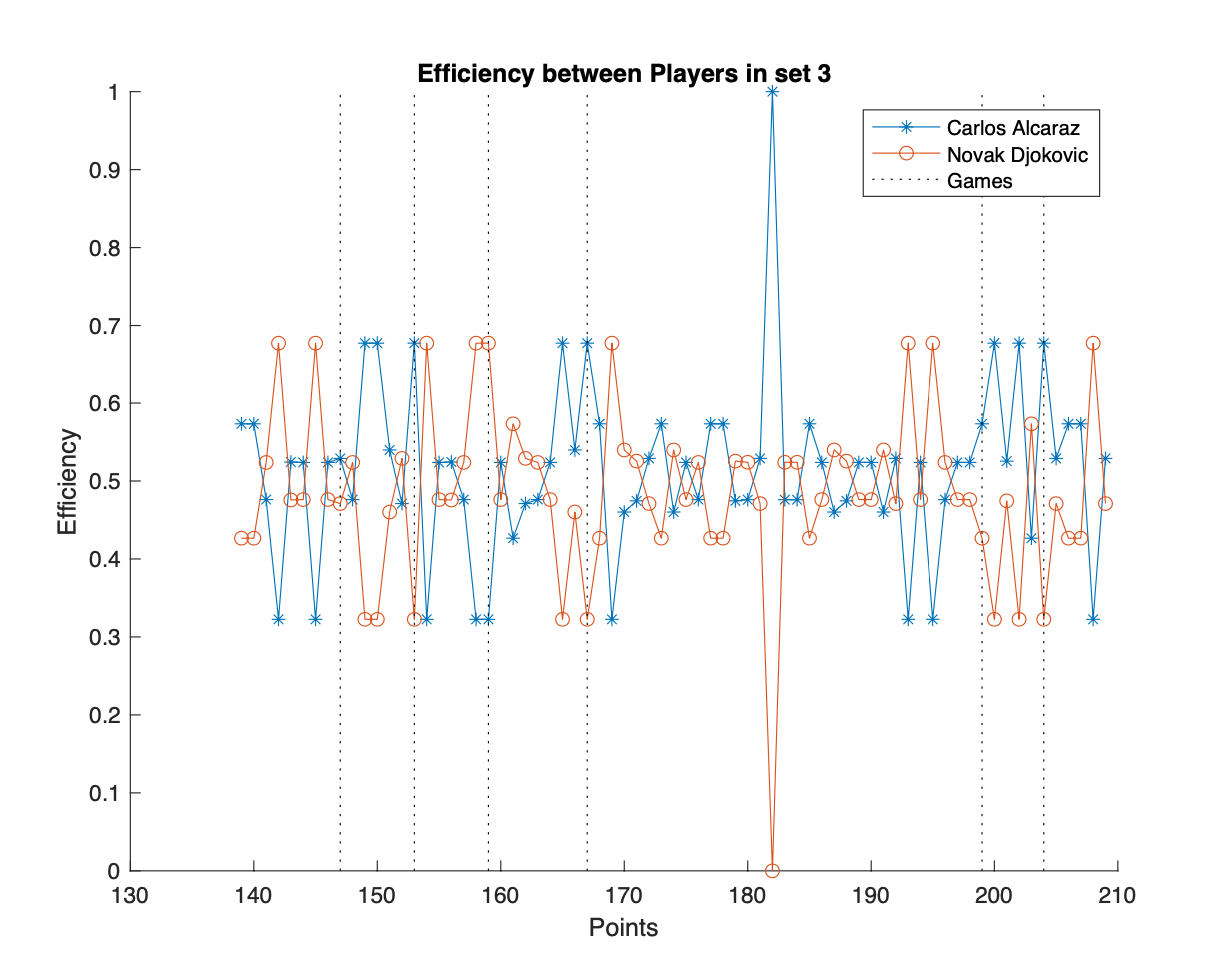}
    \caption{An example of efficiency in a single set}
\end{figure}

In conclusion, we have developed an efficiency factor $E$ that is a function of the number of rallies $k$ in a point, reflecting how quickly a player is able to win a point and how this efficiency contributes to the development of momentum. This approach allows us to capture the dynamic nature of momentum in tennis, with shorter rallies resulting in higher efficiency and longer rallies leading to lower efficiency.

Additionally, the efficiency factor can be further refined by incorporating other match dynamics, such as consecutive points won in a row and the occurrence of double faults. These factors can modify the efficiency, as consecutive points won demonstrate a player’s ability to build momentum, while double faults represent an error that decreases a player's efficiency and disrupts the flow of the match. By considering these additional factors, we can create a more comprehensive model of momentum that reflects the complexities of real match dynamics.

Ultimately, this efficiency factor forms a key component of the Tennis Momentum Model (TMM), allowing us to track how momentum shifts throughout a match and providing valuable insights into how players can leverage efficiency to influence match outcomes.

\section{Momentum Model}
\subsection{Prediction Model}
Building upon the basic model, we introduce short-term momentum that deviates from the iid assumption. Short-term momentum is defined as a temporary advantage in scoring based on recent events within the match, such as winning consecutive points or successfully executing powerful shots. Unlike the iid assumption where scoring probabilities remain constant, short-term momentum reflects the fact that momentum can shift during the match, influencing the likelihood of a player winning a point.

In our model, the short-term momentum, denoted as
$M_{stm}$, is now proportional to the historical scoring probability $p_{hist}$. Specifically, we can express this relationship as:
$$ M_{stm} = p_{hist} \dot  f(E) $$
Where:
\begin{itemize}
\item $p_{hist}$  is the historical scoring probability (i.e., the player’s average probability of winning a point based on past data).
\item $f(E)$ is a function that captures the effect of short-term momentum, where $E$ represents the efficiency. This function can be tailored to represent how recent events (such as winning consecutive points, aces, or other momentum-building actions) influence the player’s chances of winning the next point (simple example like $f(E)=E$).
\end{itemize}

In this way, short-term momentum adjusts the scoring probability, making it a more responsive and accurate reflection of the current match dynamics.

\subsection{Full Model}
 With these relations in mind a model is proposed for the probability of winning a serve that considers long-term scoring probability $p_{ltm}$
\begin{equation}
    p_{ltm} = p_{hist} - (k \times T_{points} \times p_{hist}) + (k \times T_{points} \times p_{inst}).
\end{equation}
In this model $k$ is a constant and $T_{points}$ is the amount of scored thus far in a match. Our reason for using $T_{points}$ is that it can efficiently map the progress of the match. We know $k$ must be less than 1 in order to capture the gradual change in probability. From this assumption we set $k = \frac{1}{\lambda}$ and rewrite the model
\begin{equation}
    p_{ltm} = p_{hist} - \frac{T_{points} \times p_{hist}}{\lambda} + \frac{T_{points} \times p_{inst}}{\lambda}.
\end{equation}

We set $\lambda$ to be the total estimated points of a match \cite{Volleyball}. The reasoning behind this is that as the match reaches it conclusion the best predictor of the probability will now entirely be the in match probability and not the current probability. The model for $p_{ltm}$ is now
\begin{equation*}
    p_{ltm} = p_{hist} - \frac{T_{point}}{E_{points}}p_{hist}+\frac{T_{point}}{E_{points}}p_{inst},
\end{equation*}
where $E_{points}$ is the expected points in a match. We calculate $E_{points}$ as being the sum of the expected points of the two players, and the expected points of a player is found from historical data.

Now having derived a model for the efficiency and the long-term scoring probability, these two are combined into our full TMM. We theorize that total momentum is a product of the efficiency and the long-term scoring probability. With this assumption, the TMM is proposed as
\begin{equation}
     TMM = \Bigr[ p_{hist} - \frac{T_{point}}{E_{points}}p_{hist}+\frac{T_{point}}{E_{points}}p_{inst} \Bigr] \times E,
\end{equation}
Where $TMM$ is the momentum a player has during a given point. This model for $TMM$ allows for a dynamic representation of how long-term factors in a match, such as scoring percentage and score count as well as shorter-term factors such as player efficiency are influencing the flow of a match.

\section{Experiments on Match}
Using our model, we decided to test it out on the 2023 Men's Wimbledon Final, consisting of Carlos Alcaraz and Novak Djokovic  and one regular game between Carlos Alcaraz and Daniil Medvedev \cite{data}.

\subsection{Carlos Alcaraz versus Novak Djokovic}
The 2023 Men’s Wimbledon Final between Carlos Alcaraz and Novak Djokovic was a perfect test case for our model, as it was anything but one-sided. Both players fought off momentum swings, producing an instant classic that went to five sets—an ideal scenario to assess the effectiveness of our momentum tracking model.

To test the long-term momentum model, we started by analyzing the previous four matches for both Alcaraz and Djokovic. From this historical data, we calculated key statistics, including their serve percentage, average match length, and predicted match length for the final. These factors served as inputs to our long-term momentum model, which aimed to capture how each player’s overall performance and efficiency could influence the flow of the final.

In Figure 2, the graph illustrating the long-term probability for both players can be seen. These graphs show how the long-term momentum of Alcaraz and Djokovic evolves as the match progresses. By comparing the momentum trajectories for both players, we can observe how differences in serve percentage, match length, and other factors impact the probability of winning subsequent points and games as the match continues.

This graph offers insight into how momentum shifts between the two players throughout the match, with potential turning points where one player’s momentum increases significantly over the other. Such insights are invaluable for understanding the strategic adjustments that may have taken place during the match.
\begin{figure}[h!]
    \centering
    \includegraphics[width=0.7\linewidth]{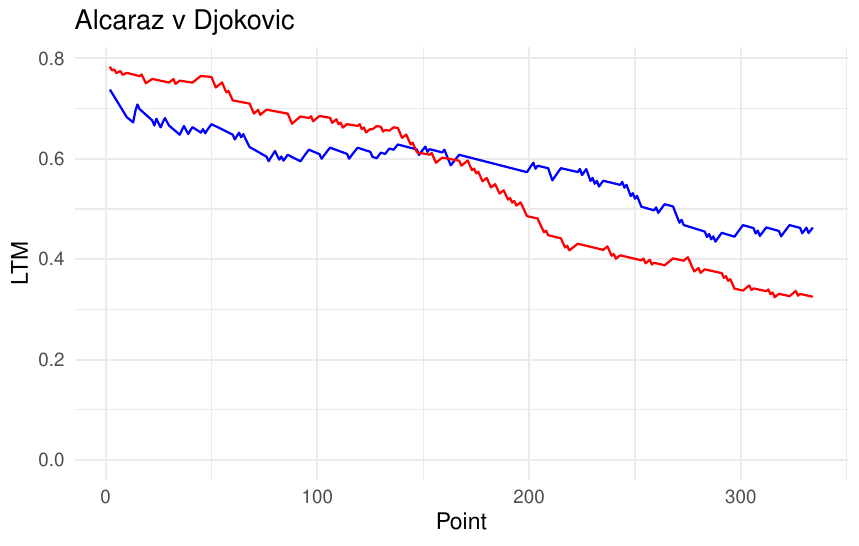}
    \caption{Alcaraz is represented as the blue line and Djokovic is represented as the red line.}
\end{figure}

The Tennis Momentum Model (TMM) accurately portrays the flow of the match between Carlos Alcaraz and Novak Djokovic. Djokovic started strong, dominating the first set, but during the second set, Alcaraz steadied the match. The momentum shift became evident just after Alcaraz won the second set, and this momentum shift propelled him to a dominant third-set victory.

Alcaraz’s success in the third set can largely be attributed to his improved serve percentage as the match progressed. Our model captured this shift in momentum, showing how Alcaraz was able to capitalize on his increased efficiency. On the other hand, in the fourth set, Djokovic began to fight back into the match, showing a gradual rise in his momentum. However, despite his efforts, Alcaraz's long-term momentum continued to build, and he ultimately sealed the victory in the fifth set.

Figure 2 illustrates the overall match probability throughout the course of the game. The figure clearly shows the momentum shifts, with Djokovic's early advantage, Alcaraz’s resurgence, and Djokovic’s brief recovery in the fourth set. However, as discussed earlier, momentum efficiency plays a crucial role in understanding these shifts.

In addition to tracking overall momentum, we further analyzed rally count efficiency throughout each point. By examining the number of rallies per point, we were able to determine which player was gaining efficiency at specific moments. Figure 3 visualizes this efficiency, providing a deeper understanding of how each player’s performance in individual rallies contributed to the larger momentum shifts in the match.
\begin{figure}[h!]
    \centering
    \includegraphics[width=0.8\linewidth]{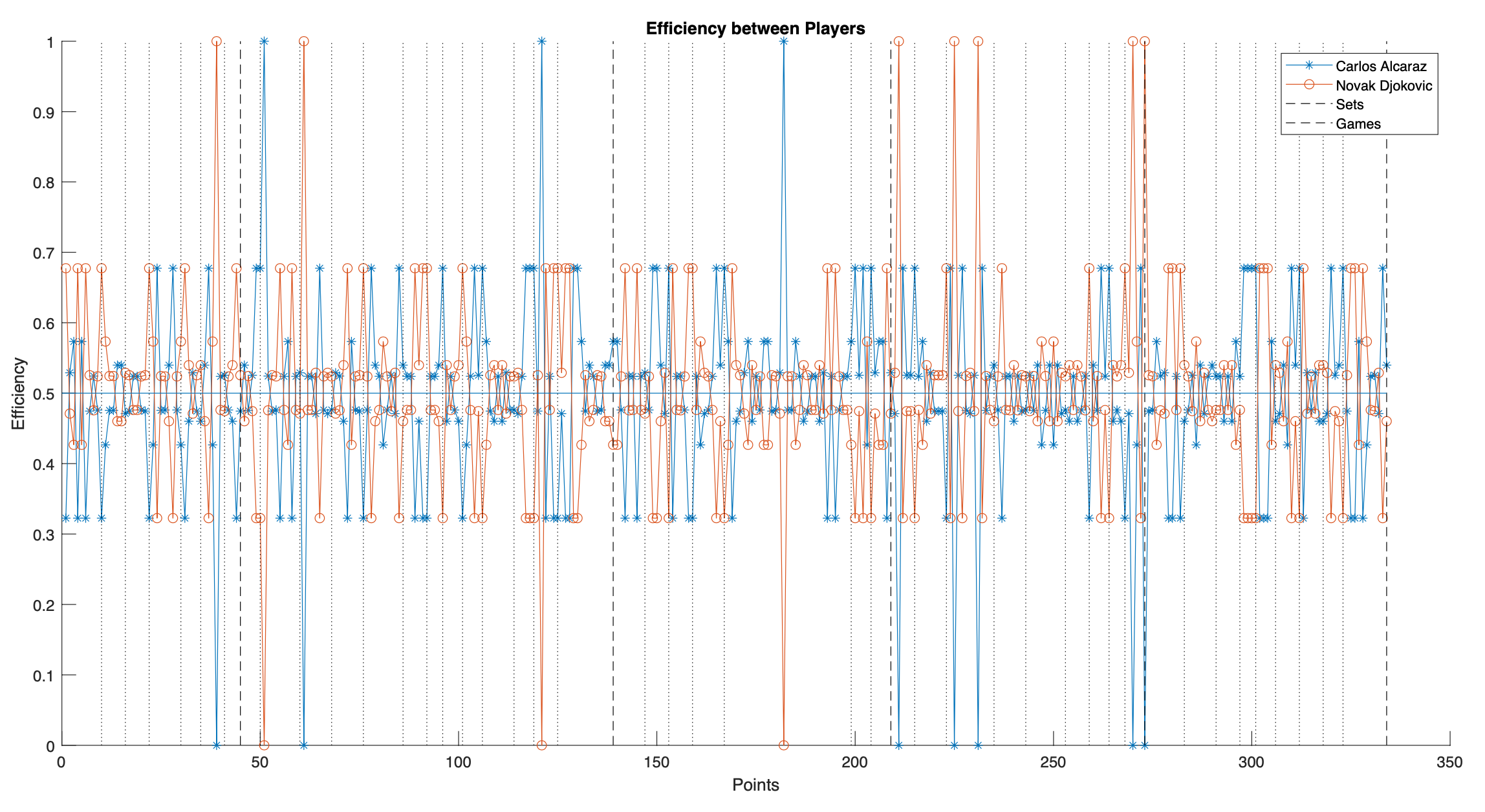}
    \caption{Alcaraz versus Djokovic efficiency.}
\end{figure}

Using the data gathered from the short-term probability, long-term probability, and efficiency metrics, we combined these components to generate a comprehensive match flow graph. Figure 4 accurately illustrates the trajectory of the match, highlighting Djokovic's early dominance. As the graph shows, Djokovic gained an early advantage in the match, but once Alcaraz regained control after winning the second set, he was able to maintain his momentum throughout the remainder of the match. This shift in momentum, as captured by the model, ultimately allowed Alcaraz to lift the trophy after a well-fought battle.

The beauty of this model lies in its real-time application. Since all the graphs are based on in-match statistics, coaches and spectators can use the momentum model during the match to assess how momentum is building for each player. By observing the live match flow graph, they can make predictions about the direction of the match—whether a player is gaining momentum or if the momentum is shifting back and forth, helping inform coaching strategies or provide deeper insights into the match dynamics.
\begin{figure}[h!]
    \centering
    \includegraphics[width=0.8\linewidth]{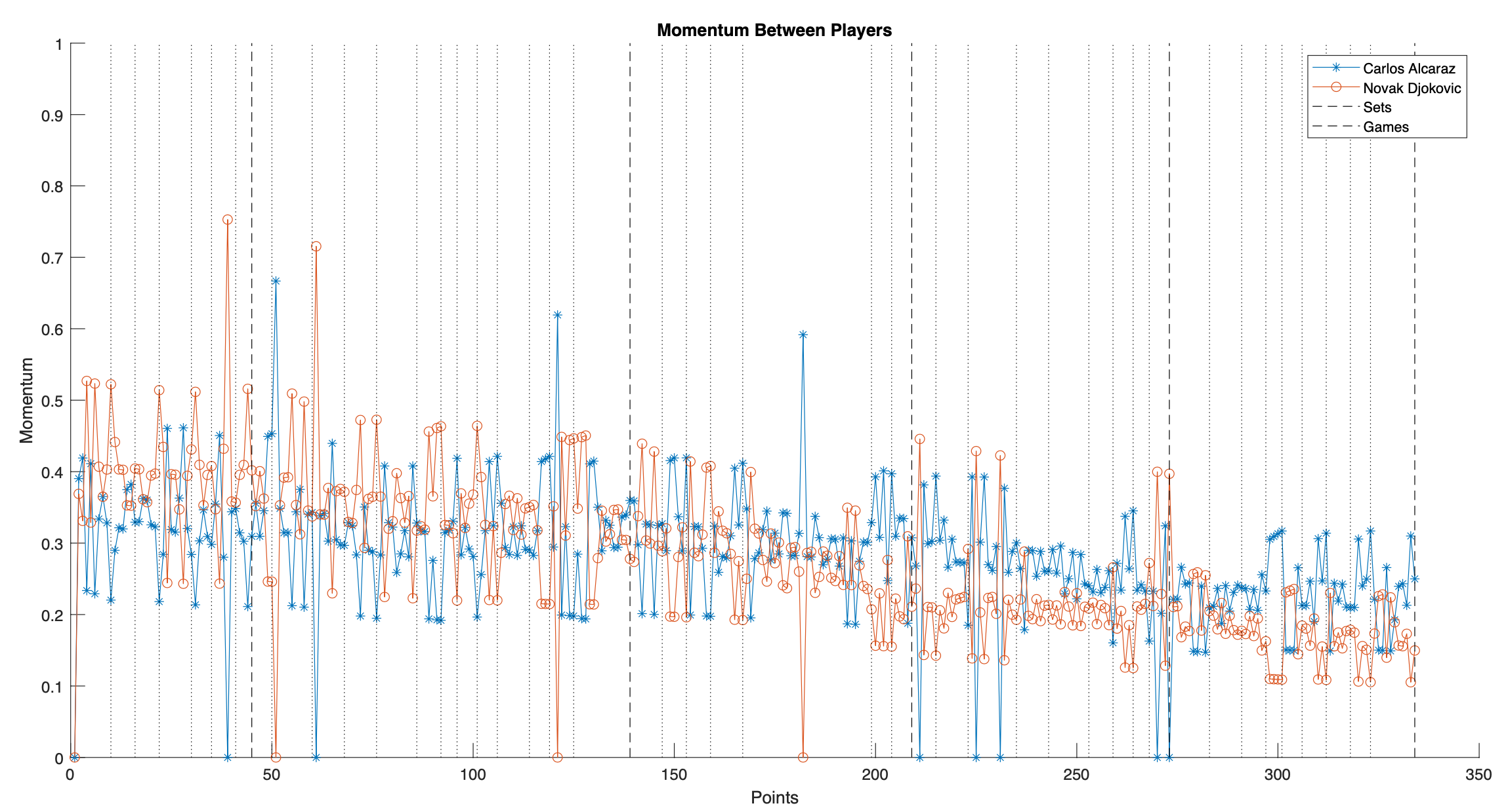}
    \caption{Alcaraz versus Djokovic complete momentum.}
\end{figure}

\subsection{Carlos Alcaraz versus Daniil Medvedev}
To further validate the accuracy and robustness of our model, we decided to test it on a one-sided match. In the 2023 Wimbledon semifinals, Carlos Alcaraz faced Daniil Medvedev in a brief encounter, where Alcaraz won all three sets 6-3, with little resistance from Medvedev. This one-sided match provided an ideal opportunity to see how our model captures a scenario where momentum remains relatively stable for one player throughout the match.

Similar to the 2023 Wimbledon Final, we calculated the long-term probability for both players by analyzing data from their previous three matches. This allowed us to generate a reliable baseline for the players’ historical performance leading up to the semifinal. The resulting long-term probability graph for this match is shown in Figure 5.

As expected, the graph for this match demonstrated a clear and consistent upward trajectory for Alcaraz, reflecting his dominance in the match. In contrast, Medvedev’s probability remained low throughout, indicating a lack of momentum shifts or major gains during the course of the match. This data highlights the model's ability to capture the disparity in momentum between the two players in a one-sided match.
\begin{figure}[h!]
    \centering
    \includegraphics[width=0.8\linewidth]{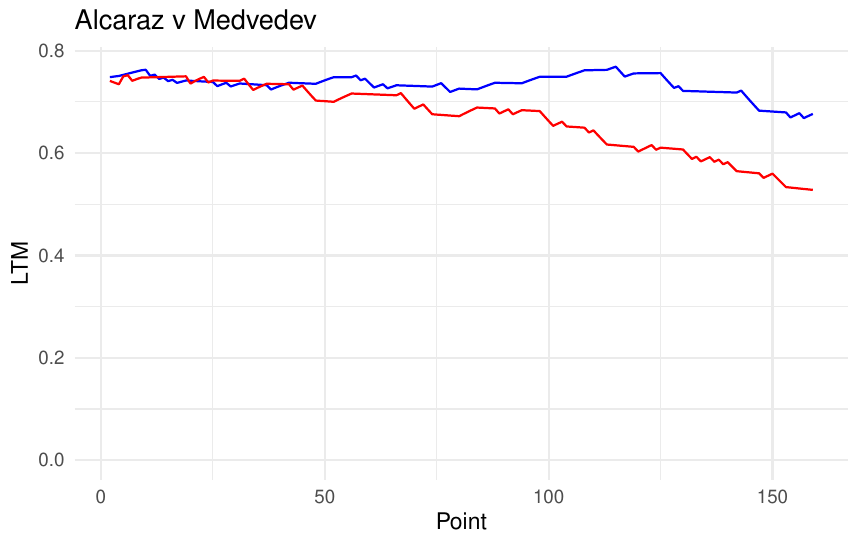}
    \caption{Alcaraz is represented as the blue line and Medvedev is represented as the red line.}
\end{figure}

At the start of the semifinal match against Daniil Medvedev, Carlos Alcaraz came out with exceptionally high long-term momentum, as indicated by the data from his previous three matches. This early momentum, combined with his high rolling serve percentage, made him look nearly unstoppable. The model accurately tracked this early surge in momentum, reflecting his commanding presence on the court from the very beginning.

As shown in Figure 6, which displays in-game efficiency, Alcaraz's performance continued to mirror this upward trend. His efficiency, which takes into account key factors such as rally count, serve performance, and point length, reinforced the idea that he was consistently outperforming Medvedev. Alcaraz’s high efficiency in rallies, combined with his strong serving, allowed him to maintain control throughout the match, leading to his dominant win.

These insights from the model highlight how long-term momentum and in-game efficiency work in tandem to give a player a significant advantage, especially in a match where momentum is largely one-sided. The ability of the model to capture this dynamic is crucial for understanding how Alcaraz was able to maintain his level of play throughout the match.
\begin{figure}[h!]
    \centering
    \includegraphics[width=0.6\linewidth]{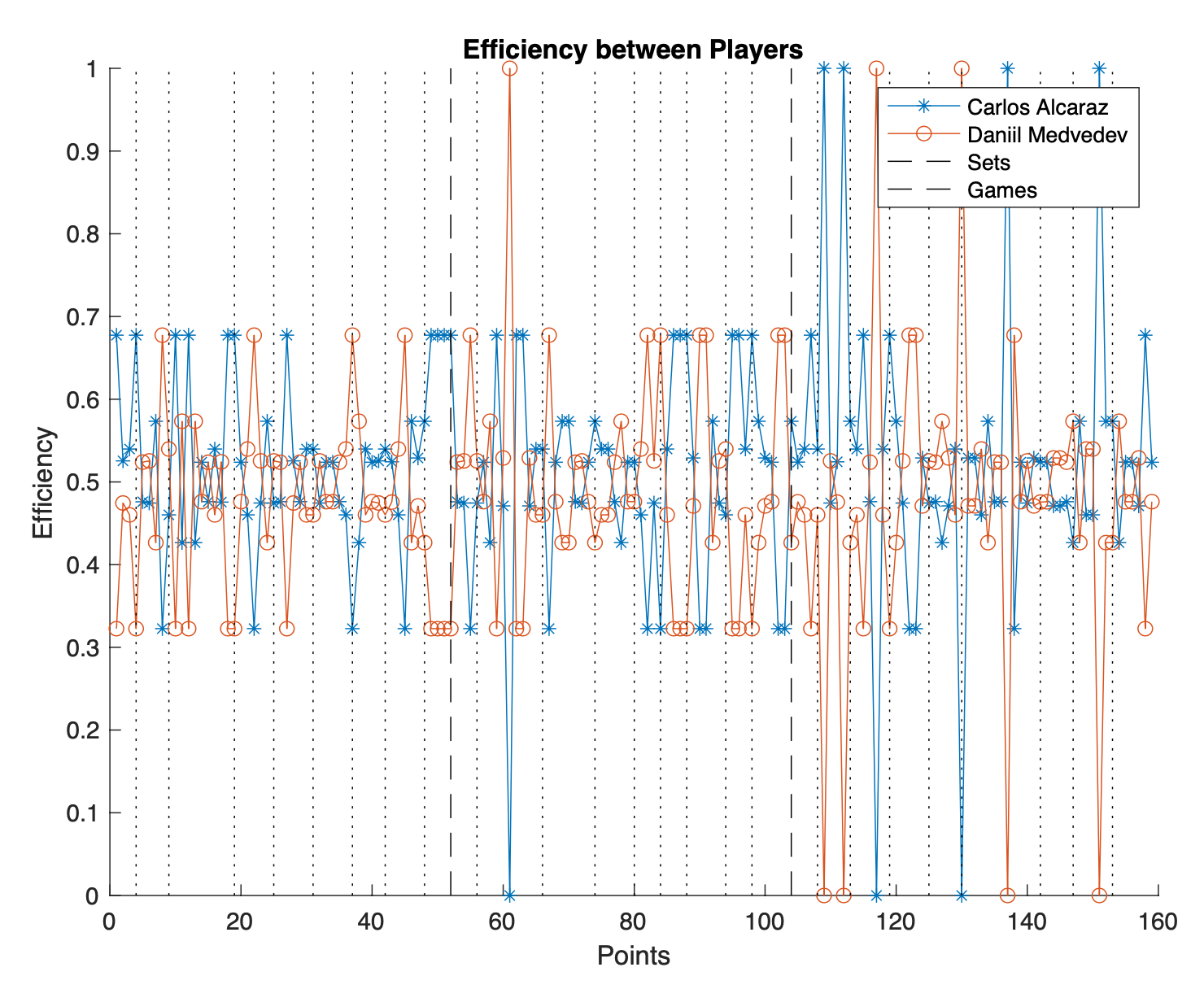}
    \caption{Alcaraz versus Medvedev efficiency.}
\end{figure}

By combining the insights from Figure 5 (long-term probability) and Figure 6 (in-game efficiency), we are able to produce a complete match flow graph that accurately represents the course of the 2023 Wimbledon semifinal. This graph shows a clear and consistent win for Carlos Alcaraz, validating the effectiveness of our model in capturing momentum shifts and player performance.

This match, in contrast to the championship match, was notably different in terms of momentum dynamics. While Alcaraz and Djokovic battled back and forth in a classic, momentum-heavy encounter in the final, Alcaraz’s semifinal match against Medvedev saw little change in momentum. From the outset, Alcaraz maintained a commanding lead, and the model accurately reflected this, with little fluctuation in momentum throughout the match.

Despite the complete control that Alcaraz exhibited, the model was able to adjust to the game flow, correctly predicting the match’s trajectory and confirming Alcaraz’s dominant performance. The absence of significant momentum swings in this one-sided match presented a unique test for the model, yet it still produced a reliable and accurate representation of the match dynamics, validating the model's robustness in both competitive and one-sided match scenarios.

\begin{figure}[h!]
    \centering
    \includegraphics[width=0.75\linewidth]{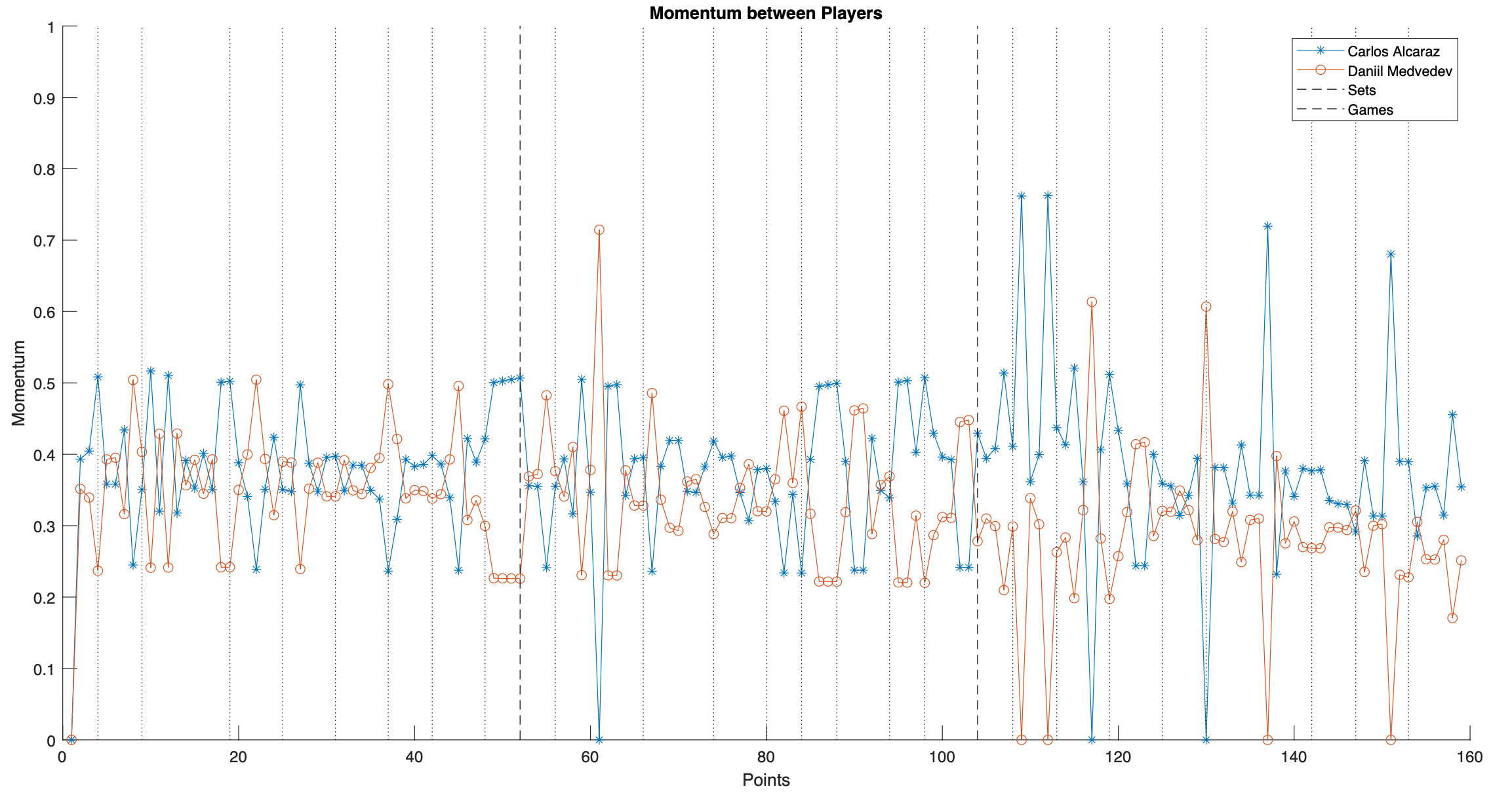}
    \caption{Alcaraz versus Medvedev complete momentum.}
\end{figure}

\section{Flaws}
One of the key advantages of our model is its use of recent data from the Wimbledon event, which provides up-to-date player statistics. This allows for more accurate estimations of expected win percentages and expected points within a match, enhancing the model's efficiency. The ability to calculate historical statistics also helps fine-tune the model and improve its predictions. While these data points are not strictly necessary, they certainly contribute to enhancing the model's performance and overall accuracy.

However, there are a few limitations to our model that we have considered. One major concern is the limited scope of variables incorporated into the model. For example, we do not account for the psychological momentum driven by the environment—such as the crowd’s influence—on player performance. A roaring crowd can significantly impact a player's psychological state, potentially shifting their momentum. This is something our current model does not capture.

Additionally, we have not considered factors such as preexisting injuries, which can drastically affect a player's performance during a match. While these external influences are important, they are currently outside the scope of our model. The absence of these variables may limit the model's accuracy in real-world scenarios, particularly in matches where psychological factors or physical limitations play a significant role.

Another consideration is the assumption that player efficiency—which is loosely defined as how few rallies a player needs to win a point—is positively correlated with momentum. This assumption generalizes tennis players’ tendencies, suggesting that all players perform better with shorter rallies. However, this might not always be the case. For example, certain players may thrive in longer rallies and actually gain more momentum from extended exchanges. Our model currently does not account for this player preference, which could skew momentum estimations for certain players.

This is by no means a comprehensive list of limitations, but these are some of the key concerns we've identified so far. Perhaps the most pressing issue is the lack of extensive testing of our model. To address this, we are considering employing a Monte Carlo simulation for future testing. By running multiple simulations, we can assess the model's ability to predict match outcomes with greater accuracy. Moreover, we could experiment by treating the efficiency and short-term momentum factors as random variables and compare the results with the current method. This would provide valuable insights into whether our current approach to modeling short-term momentum is effective in predicting momentum shifts in real matches.

\section{Conclusion}
In this paper, we tackled the challenge of quantifying momentum in tennis matches. Through testing and results, we demonstrated that momentum plays a significant role in the game, leading us to develop the Tennis Momentum Model (TMM). This model combines both short-term and long-term momentum to accurately capture the dynamic flow of a match as it unfolds. Coaches and players can leverage the TMM to enhance their understanding of momentum, providing them with insights to adjust strategies and improve performance during the match.

The TMM successfully represents the varying momentum between players, making it a powerful tool for both coaches and players. By incorporating key factors such as rally count, serve percentage, aces, small samples of historical data, and real-time match outcomes, the model effectively tracks momentum shifts. The probability and efficiency functions in the model allow for a comprehensive understanding of how momentum changes throughout a match. This can highlight areas where players may need to improve, whether it be serving, rally management, or other aspects of their game.

With additional historical data and the integration of a random data simulator, the TMM can be further refined to predict potential outcomes of future matches. This would allow coaches to interpret momentum models based on past performances and apply insights to specific matchups, whether against the same opponent or different ones. Coaches can then focus on improving particular aspects of their player's game, such as breaking an opponent's momentum by adjusting their return game, whether through techniques like forehands or backhands.

Moreover, by analyzing the opponent's momentum model, coaches can identify weaknesses in their opponent's game, giving their players a competitive edge. This ability to anticipate and respond to momentum shifts, based on both their own player's performance and their opponent's tendencies, would significantly improve a coach's ability to influence match outcomes.

However, it is important to acknowledge the limitations of the model. Tennis is a dynamic, unpredictable sport, and many external factors—such as environmental influences, psychological state, and personal circumstances—are not captured in the model. For instance, a player's pre-match mindset, injuries, or the impact of the crowd can all influence their performance in ways the TMM cannot predict. These variables remain outside the scope of the model but should be considered by coaches who use it. A key part of the model’s effectiveness lies in the ability of coaches to adapt it in real-time based on their observations and knowledge of the players’ mental and physical states.

Additionally, random events, such as comments from opponents, crowd energy, or unplanned incidents, can skew the momentum flow in unpredictable ways. These external elements will always introduce a level of error, but the TMM, when combined with the practical knowledge of coaches, can still provide valuable insights into match dynamics.

In conclusion, while no model can perfectly predict momentum shifts or match outcomes, the Tennis Momentum Model (TMM) is a powerful tool for understanding the flow of a tennis match. It offers coaches and players valuable insights into how momentum is shifting and where adjustments can be made to improve performance. By combining the TMM with a coach’s expertise, players can gain a significant advantage, helping them win more matches and perform at their best.

\bibliographystyle{plain}

\end{document}